\documentclass[twocolumn,aps,superscriptaddress,showpacs,nofootinbib]{revtex4}

\usepackage{amssymb}
\usepackage{amsmath}
\usepackage{graphicx}
\usepackage[normalem]{ulem}
\usepackage[dvips]{color}

\usepackage[normalem]{ulem}  
\usepackage[dvips]{color} 

\renewcommand\sout{\bgroup \color{red} \ULdepth=-.5ex \ULset}

\begin{document}
\title{Collision energy dependence of elliptic flow splitting between particles and their antiparticles from an extended multiphase transport model}

\author{Jun Xu}
\affiliation{Shanghai Institute of Applied
Physics, Chinese Academy of Sciences, Shanghai 201800, China}

\author{Che Ming Ko}
\affiliation{Cyclotron Institute and
Department of Physics and Astronomy, Texas A$\&$M University,
College Station, Texas 77843, USA}

\date{\today}

\begin{abstract}
Based on an extended multiphase transport model, which includes
mean-field potentials in both the partonic and hadronic phases, uses the mix-event coalescence,
and respects charge conservation during the hadronic evolution, we have studied the collision energy
dependence of the elliptic flow splitting between particles and
their antiparticles. This extended transport model reproduces reasonably well the
experimental data at lower collision energies but only describes
qualitatively the elliptic flow splitting at
higher beam energies. The present study thus indicates the existence of other mechanisms 
for the elliptic flow splitting besides the mean-field potentials and the need of further improvements of the multiphase
transport model.
\end{abstract}


\maketitle

\section{introduction}
\label{introduction}

Understanding the phase diagram of quantum chromodynamics (QCD) is
one of the main goals of heavy-ion collision experiments. From
lattice QCD calculations, it is found that the hadron-quark phase transition is
a smooth crossover at zero baryon chemical
potential~\cite{Ber05,Aok06,Baz12}, corresponding to the matter at the top energy of relativistic heavy-ion collider (RHIC)
and the large hadron collider (LHC). At finite baryon chemical potential
as expected in lower collision energies, various theoretical models, e.g., the
Nambu-Jona-Lasinio (NJL) model~\cite{Asa89,Fuk08,Car10,Bra13}, have shown that the hadron-quark phase transition can be a first-order one. In
order to search for the signal of the QCD critical point, which lies
at the phase boundary between the smooth crossover transition to the
first-order transition, great efforts have been made in several
experimental programs. Among them, the Phase I of the beam-energy
scan (BES) program has been carried out, and the Phase II of BES
program as well as the compressed baryonic matter (CBM) program at
the international Facility for Antiprotons and Ion Research (FAIR)
are being planned.

The RHIC-BES I has
led to many exciting findings, and among them the elliptic flow ($v_2$) splitting between
particles and their antiparticles has attracted considerable
theoretical attentions. For example, the $v_2$ splitting can be
attributed to the larger $v_2$ for transported quarks than that for
produced quarks, or similarly, their different rapidity
dependencies~\cite{Dun11,Gre12,Iva13}, hydrodynamic evolution of the
QGP at finite baryon chemical potential~\cite{Ste12,Hat15}, and
the smaller radial flow of particles than their
antiparticles~\cite{Sun15}. The $v_2$ splitting between $\pi^-$ and
$\pi^+$ can also be attributed to the electric quadrupole moment
in the produced quark-gluon plasma (QGP) due to the effect from the
chiral magnetic wave~\cite{Bur11} and their similar dynamics in the hadronic matter.

In our previous studies~\cite{Xu14} by introducing mean-field
potentials in both the partonic phase~\cite{Son12} and the hadronic
phase~\cite{Xu12} of a multiphase transport (AMPT) model, we
reproduced the relative elliptic flow splitting between protons and
antiprotons as well as kaons and antikaons at only the lowest energy
of RHIC-BES, i.e., 7.7 GeV. Based on our previous models, we have
made further modifications of the AMPT model in addition to the
incorporation of the mean-field potential, such as applying the
mix-event coalescence algorithm in the hadronization process and
correcting the charge conservation in the hadronic phase.
In the present study, we use this extended AMPT model to
explore the collision energy dependence of elliptic flow splitting between particles and their antiparticles in heavy ion collisions.

This paper is organized as follows. In Sec. \ref{potential}, we review the partonic mean-field potentials based on a 3-flavor NJL model and the hadronic mean-field potentials from effective Lagrangians. The extensions made on the original AMPT model as well as the structure of the extended AMPT model are briefly described in Sec. \ref{ampt}. The dynamics of produced baryon-rich matter and detailed results on the elliptic flow splitting between nucleons and antinucleons as well as that between $K^+$ and $K^-$ at various collision energies are discussed in Sec. \ref{results}. In Sec.\ref{summary}, we summarize our results and discuss possible extensions of the present model and its applications.

\section{Mean-field potentials}\label{potential}

\subsection{Partonic mean-field potentials}

We first briefly review the mean-field potentials for particles
and their antiparticles in the partonic and hadronic phase as used
in our previous studies~\cite{Xu14}. The partonic mean-field potentials in the baryon-rich quark matter are calculated from a 3-flavor NJL model~\cite{Son12} with a Lagrangian given by~\cite{Bra13}
\begin{eqnarray}
\mathcal{L_{NJL}}&=&\bar{q}(i\not{\partial}-M)q+\frac{G}{2}\sum_{a=0}^{8}\bigg[(\bar{q}\lambda^aq)^2+(\bar{q}i\gamma_5\lambda^aq)^2\bigg]\nonumber\\
&+&\sum_{a=0}^{8}\bigg[\frac{G_V}{2}(\bar{q}\gamma_\mu\lambda^aq)^2+\frac{G_A}{2}(\bar{q}\gamma_\mu\gamma_5\lambda^aq)^2\bigg]\nonumber\\
&-&K\bigg[{\rm det}_f\bigg(\bar{q}(1+\gamma_5)q\bigg)+{\rm
det}_f\bigg(\bar{q}(1-\gamma_5)q\bigg)\bigg],
\end{eqnarray}
where $q=(u, d, s)^T$ is the quark field,
$M={\rm diag}(m_u, m_d, m_s)$ is the current quark mass matirx, and
$\lambda^{a}$ is the Gell-Mann matrices in $SU(3)$ flavor space with $\lambda^0=\sqrt{2/3}I$.
In the case that the vector and axial-vector interactions are
generated by the Fierz transformation of the scalar and
pseudo-scalar interactions, their coupling strengths are given by
$G_V=G_A=G/2$, while $G_V=1.1G$ was used in Ref.~\cite{Lut92} to
give a better description of the vector meson-mass spectrum based on
the NJL model. The last term in Eq.(1), with the det$_f$ denoting the determinant in the flavor space, is the Kobayashi-Maskawa-t'Hooft (KMT) interaction~\cite{Hoo76} that breaks the axial $U(1)_A$ symmetry.

In the mean-field approximation, the quark effective masses are given by
\begin{eqnarray}
M_u &=& m_u - 2G\langle \bar{u}u \rangle + 2K\langle \bar{d}d \rangle \langle \bar{s}s \rangle = m_u + \Sigma_s^u,\label{mu}\\
M_d &=& m_d - 2G\langle \bar{d}d \rangle + 2K\langle \bar{s}s \rangle \langle \bar{u}u \rangle = m_d + \Sigma_s^d,\label{md}\\
M_s &=& m_s - 2G\langle \bar{s}s \rangle + 2K\langle \bar{u}u \rangle \langle \bar{d}d \rangle = m_s + \Sigma_s^s,\label{ms}
\end{eqnarray}
where the quark condensate is
\begin{eqnarray}
\langle \bar{q_i}q_i \rangle &=& -2 M_i N_c \int \frac{d^3k}{(2\pi)^3E_i}[1-f_i(k)-\bar{f}_i(k)],\nonumber\\
&&(i=u, d, s)\label{qq}
\end{eqnarray}
with the number of colors $N_c=3$, the single-quark energy $E_i=\sqrt{M_i^2+k^2}$, and $f_i(k)$ and $\bar{f}_i(k)$ being the phase-space distribution functions of quarks of flavor $i$ and its anti-flavor, respectively. An iteration method is needed to calculate the effective mass $M_i$ and the scalar potential $\Sigma_s^i$ of flavor species $i$ from Eqs.~(\ref{mu}), (\ref{md}), (\ref{ms}), and (\ref{qq}).

From the flavor-average treatment employed in
Refs.~\cite{Asa89,Hat94}, the vector part in the Lagrangian is taken
as $g_V\langle \bar{q} \gamma_\mu q \rangle^2$ with $g_V=(2/3)G_V$,
and in this way the single-particle Hamiltonian of quark flavor $i$
with momentum $\vec{p}$ is written as
\begin{equation}
H_i = \sqrt{M_i^2+(\vec{p} \mp g_V \vec{\rho})^2} \pm g_V \rho^0,
\end{equation}
where
\begin{equation}
\rho^\mu = 2N_c \sum_{i=u,d,s} \int \frac{d^3k}{(2\pi)^3E_i} k^\mu [f_i(k)-\bar{f}_i(k)] \label{rhov}
\end{equation}
is the vector density with $\rho^0$ being its time component, i.e.,
the net quark density. As discussed in Ref.~\cite{Son12}, the time
component of the vector potential $\Sigma_v^0=g_V\rho^0$ is more
important than its space component in heavy ion collisions at 7.7 GeV. The reason is that
the space component of the vector potential is related to the
current that needs time to develop, while the elliptic flow is
mostly produced at the early stage of the partonic phase.

\begin{figure}[h]
\centerline{\includegraphics[scale=0.8]{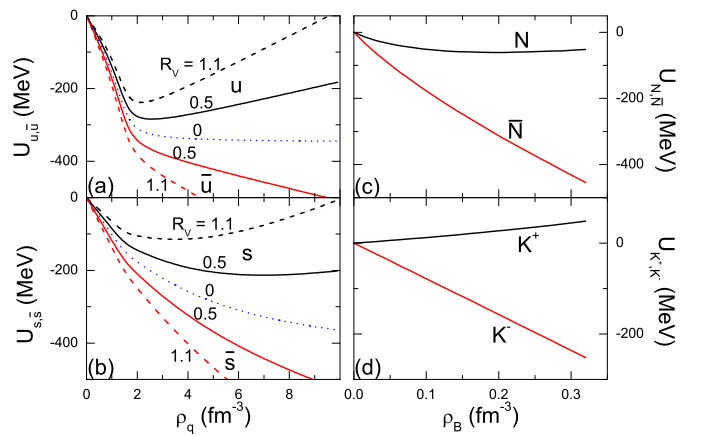}} \caption{(Color
online) Reduced non-relativistic mean-field potentials for up quarks
and anti-up quarks (a), strange quarks and anti-strange quarks (b),
nucleons and antinucleons (c), and $K^+$ and $K^-$ (d) in the cold antiquark- or antibaryon-free nuclear medium.} \label{U}
\end{figure}

As the NJL model is not renormalizable, the momentum integrations in
Eqs.~(\ref{qq}) and (\ref{rhov}) require a cut-off momentum $\Lambda$. Taking
$\Lambda=750$ MeV~\cite{Lut92,Bra13}
and the current quark masses $m_u=m_d=3.6$ MeV and $m_s=87$ MeV, the values $G$ and $K$ can be
determined from fitting the pion and kaon masses as well as the pion decay constant, and
their values are $G\Lambda^2=3.6$ and
$K\Lambda^{5}=8.9$~\cite{Lut92,Bra13}. Although the dynamics of partonic matter is
treated relativistically in transport simulations, it is instructive
to show the non-relativistic reduction of the mean-field potential
$U_{q_i,\bar{q_i}}=\Sigma_s^i \pm \Sigma_v^0-M_c^i$, where $M_c^i$
is the constituent quark mass in vacuum. As an illustration, this
potential is shown in panel (a) of Fig.~\ref{U} for $u$ and
$\bar{u}$ as well as in panel (b) of Fig.~\ref{U} for $s$ and
$\bar{s}$ in a quark matter with equal density for $u$, $d$, and $s$
quarks at zero temperature for the cases of $R_V=G_V/G=0$, $0.5$,
and $1.1$. The mean-field potentials for $d$ and $\bar{d}$ are
exactly the same as those for $u$ and $\bar{u}$ as we have not
included isovector coupling in the NJL model. Although the scalar
potential $\Sigma_s^i$ for both quarks and antiquarks is attractive
after subtracting $M_c^i$ (see the curve with $R_V=0$), the vector
potential is repulsive for quarks and attractive for antiquarks, and
this makes the potential for antiquarks more attractive than that
for quarks. As seen in Fig.~\ref{U}, the potential difference
between quarks and antiquarks increases with increasing quark
density $\rho_q$ and increasing value of $R_V$. Because of the sufficiently
large value of $\Lambda$ used in our study, results presented in the following
are not expected to be sensitive to its exact value.

\subsection{Hadronic mean-field potentials}

For the nucleon and antinucleon potentials, we use those from the
relativistic mean-field theory based on the following Lagrangian~\cite{GQL94}:
\begin{eqnarray}\label{lag}
\mathcal{L_H} &=&
\overline{\psi}[i\gamma_\mu\partial^\mu-m-g_\sigma\sigma-g_\omega\gamma_\mu\omega^\mu]\psi
+ \frac{1}{2}(\partial^\mu \sigma)^2 \notag\\
&-& \frac{1}{2} m_\sigma^2\sigma^2 - \frac{1}{2}b\sigma^3 -
\frac{1}{4}c\sigma^4 -
\frac{1}{4}(\partial_\mu\omega^\nu-\partial_\nu\omega^\mu)^2
\notag\\
&+& \frac{1}{2}m_\omega^2{\omega^\mu}^2,
\end{eqnarray}
where $\psi$ is the nucleon field with mass $m$, and $\sigma$ and
$\omega$ are the scalar and vector meson fields with masses
$m_\sigma$ and $m_{\omega}$, respectively. The nucleon effective mass
$m^\star$ and the kinetic momentum $p_\mu^\star$ are defined by
\begin{eqnarray}
m^\star &=& m - \Sigma_s, \label{mstar}\\
p_\mu^\star &=& p_\mu - \Sigma_{v\mu},
\end{eqnarray}
where the nucleon scalar and vector self-energies are given, respectively, by
\begin{eqnarray}
\Sigma_s &=& g_\sigma \langle \sigma \rangle, \label{sigmas}\\
\Sigma_{v\mu} &=& g_\omega \langle \omega_\mu \rangle.
\end{eqnarray}
In the mean-field approximation, the expectation values of the
scalar and vector fields in nuclear medium are related to the
nuclear scalar density $\rho_s$ and current density $\rho_\mu$ via
\begin{eqnarray}
&&m_\sigma \langle \sigma \rangle + b \langle \sigma \rangle^2 + c
\langle \sigma \rangle^3 = g_\sigma \rho_s, \label{sigmasrhos}\\
&&\langle \omega_\mu \rangle = (g_\omega/m_\omega^2) \rho_\mu.
\end{eqnarray}
In the local-density approximation, the scalar and vector densities
can be further expressed in terms of the phase-space distribution
functions $f(r,p^\star)$ of nucleons and $\bar{f}(r,p^\star)$ of antinucleons as
\begin{eqnarray}
\rho_s &=& 4\int \frac{d^3p^\star}{(2\pi)^3} \frac{m^\star}{E^\star} [f(r,p^\star)+\bar{f}(r,p^\star)]
, \label{rhos}\\
\rho_\mu &=& 4\int \frac{d^3p^\star}{(2\pi)^3} \frac{p_\mu^\star}{E^\star}
[f(r,p^\star)-\bar{f}(r,p^\star)],\label{rhomu}\\\notag
\end{eqnarray}
respectively, where $E^\star = \sqrt{{m^\star}^2+{p^\star}^2}$ is the single-particle energy. The time component of
the vector density is thus exactly the net nucleon density. The
scalar self-energy can be calculated self-consistently from
Eqs.~(\ref{mstar}), (\ref{sigmas}), (\ref{sigmasrhos}), and
(\ref{rhos}) using the iteration method. The parameters for a soft
equation of state are used in the present study~\cite{GQL94}, i.e.,
$(g_\sigma/m_\sigma)m=13.95$, $(g_\omega/m_\omega)m=8.498$,
$b/(g_\sigma^3m)=0.0199$, and $c/g_\sigma^4=-0.00296$.

Based on the G-parity invariance, the non-relativistic reduction of
the potentials for nucleons and antinucleons are
\begin{equation}
U_{N,\bar{N}} =  - \Sigma_s \pm \Sigma_v^0,\\
\end{equation}
where $\Sigma_v^0$ is the time component of the vector potential,
and the '$+$' and '$-$' signs are for nucleons and antinucleons,
respectively. As an illustration, we show in panel (c) of
Fig.~\ref{U} the potentials for nucleons and antinucleons at zero
temperature in antibaryon-free hadronic matter. It is seen that the
potential for nucleons is slightly attractive, while that for
antinucleons is deeply attractive, with the former about $-60$ MeV
and the latter about $-260$ MeV at the saturation density $\rho_0=0.16$
fm$^{-3}$. In the hadronic matter with strange baryons and baryon
resonances, the phase-space distribution functions $f(r,p^\star)$
and $\bar{f}(r,p^\star)$ are calculated from all the baryons and
antibaryons according to their light quark content.

For the kaon and antikaon potentials in nuclear medium, they are obtained from the chiral
effective Lagrangian~\cite{GQL97}, that is,
$U_{K,{\bar K}} = \omega_{K,{\bar K}} - \omega_0$ with
\begin{eqnarray}
\omega_{K,{\bar K}} &=& \sqrt{m_K^2 + p^2 - a_{K,{\bar K}}\rho_s
+(b_K\rho_B^{\rm net})^2}\pm b_K\rho_B^{\rm net}\nonumber\\
\end{eqnarray}
and $\omega_0=\sqrt{m_K^2+p^2}$, where $m_K=498$ MeV is the kaon
mass, and the values of other parameters are set as $a_K=0.22$
GeV$^2$fm$^3$, $a_{\bar K}=0.45$ GeV$^2$fm$^3$, and $b_K=0.333$
GeVfm$^3$~\cite{GQL97}. In the above, $\rho_s$ is the scalar density
determined from the effective Lagrangian in Eq.~(\ref{lag}), and
$\rho_B^{\rm net}=\rho_B - \rho_{\bar B}$ is the net baryon density.
The "$+$" and "$-$" signs are for kaons and antikaons, respectively.
The potentials for $K^+$ and $K^-$ at rest in nuclear matter at zero temperature are displayed in panel
(d) of Fig.~\ref{U}. The potential for $K^+$ is seen to be
slightly repulsive while that for $K^-$ is deeply attractive, with
the former about $20$ MeV and the latter  about $-125$ MeV at the
saturation density.

We have also introduced the $s$-wave pion potentials in the hadronic phase
as in Ref.~\cite{Xu12}. In the absence of the isovector coupling in
the partonic phase, it has been shown that the $v_2$ splitting of
$\pi^-$ and $\pi^+$ due to their potentials in the hadronic phase has the correct sign compared with the
experimental data but the magnitude is much smaller. In the present study, we thus only
discuss the $v_2$ splitting between nucleons as well as kaons and their
antiparticles, and postpone the study of the effects of isovector mean fields
in relativistic heavy-ion collisions.

\section{The extended AMPT model}\label{ampt}

To include the mean-field potentials for both partons and hadrons in
relativistic heavy-ion collisions, we have made extensive
modifications to the string melting version of the
AMPT model~\cite{Lin05,Zhang:1999bd,Lin:2001zk}. The original string melting version has
been successfully used to describe the charge particle multiplicity,
the collective flow, and the dihadron correlations in heavy-ion
collisions at the top energies at RHIC and
LHC~\cite{Xu11a,Xu11b,Xu11c,Xu11d}, where the mean-field potentials
for particles and their antiparticles are not included in either the
partonic phase or the hadronic phase as their effects are less
important than partonic and hadronic scatterings on the collision
dynamics.  However, the mean-field effects become non-negligible
in heavy-ion collisions at energies of the RHIC-BES
program and the future FAIR-CBM program. In the following, we
briefly discuss the extended AMPT model used in the present study.

The initial condition of the AMPT model is obtained from the
heavy-ion jet interaction generator (HIJING) model~\cite{Xnw91},
where both soft and hard parton production are included by using
the Monte Carlo Glauber model with shadowing effects included for
nucleus-nucleus collisions. In the original string melting version,
which converts hadrons produced from initial collisions into their
valence quarks and antiquarks, the interaction in the partonic phase
is described only by parton-parton elastic scatterings based on
Zhang's parton cascade (ZPC) model~\cite{Zha98} without mean-field
potentials for partons. In the present study as well as those
reported in Refs.~\cite{Son12,Xu14}, the ZPC model is replaced by a
3-flavor NJL transport model that includes both scalar and vector
potentials for partons as well as the parton elastic scattering
process. To calculate the mean-field potentials, the test-particle
method~\cite{Won82} with parallel events for the same impact
parameter is used. For the parton scattering cross section, it can in principle be obtained also from the NJL model~\cite{Mar13}.  In this case, the parton scattering cross section would depend on the temperature and quark chemical potential of the partonic matter. Since it is essential for our model to reproduce the experimentally measured charged particle elliptic flow before addressing the effect of mean-field potentials on the elliptical flow splitting between particles and their antiparticles, we take the value of the scattering
cross section between partons in the same event as a parameter and determine its value by fitting 
the experimental charged particle elliptic flow, as will be shown in the next section.

The partonic
evolution ends when the chiral phase transition happens, i.e., the
effective mass of light quarks in central cells of the system is
half of that in vacuum.  A spatial coalescence model as used in the original AMPT model is then used to describe the
hadronization process with the hadron species determined by the
flavor and invariant mass of its constituent quarks or antiquarks. However, instead of coalescence of quarks in the same event,
we extend the coalescence algorithm
to allow quarks and antiquarks in an event to coalesce with those in other parallel events.
The hadronization treatment of completely mixing the quarks and antiquarks from all parallel events is
equivalent to the use of smooth quark and antiquark phase-space distributions
for hadron production via quark coalescence
in the pioneering studies in Refs.~\cite{Greco:2003mm,Greco:2003xt,Fries:2003vb,Hwa:2002tu}, and is
particularly useful for rare antiparticles produced at lower collision energies.
To keep the fluctuation in the number of hadrons produced from these parallel events,
we allow, however, quarks and antiquarks in a given event to coalesce only with certain quarks and antiquarks in the parallel events so that the numbers of mesons and baryons produced in each event are the same as in the original AMPT model.
This is possible because in the string melting version of AMPT, partons in each event are obtained from converting the baryons and mesons produced from HIJING into its constituents, and daughter partons from the same hadron are labeled. In the coalescence algorithm for the hadronization of the partonic matter after its evolution, a parton recombines with other partons that are the closest in coordinate space. If the latter are originally from a different hadron, the sibling partons associated with the hadronized parton are then relabeled as the siblings of the parton whose siblings are involved in this particular coalescence. Allowing parton relabelings between parallel events thus does not alter the number of hadrons produced in a given event even the coalescence is done with mixed events. We note that although the event-by-event particle number fluctuation is maintained 
in the mixed-event coalescence method, the event-by-event density fluctuation is largely averaged out by using partons from parallel events to evaluate the mean-field potentials. The only remaining density fluctuation effect comes from the parton scatterings because only partons in the same event can scatter in this extended AMPT model.

After hadronization, a relativistic transport (ART) model is used to
describe the evolution of the hadronic phase~\cite{Bal95}, in which
both particle-antiparticle annihilations and their inverse processes
are included. The mean-field potentials for hadrons in the ART model
are also turned on~\cite{Xu12} by using the test-particle method
with parallel events. Since charges are not strictly conserved in
some of the inelastic processes in the original ART model, we have
corrected this problem by resampling the inelastic channels until
the charge is conserved.

\section{Results and discussions}\label{results}

Based on the above extended AMPT model, we have studied heavy-ion collisions at RHIC-BES energies.
Here, we focus on the evolution of the baryon-rich matter produced in these collisions, and discuss the
difference in the elliptic flows between particles and their antiparticles.

\subsection{Charged particle elliptic flow}

\begin{figure}[h]
\centerline{\includegraphics[scale=0.8]{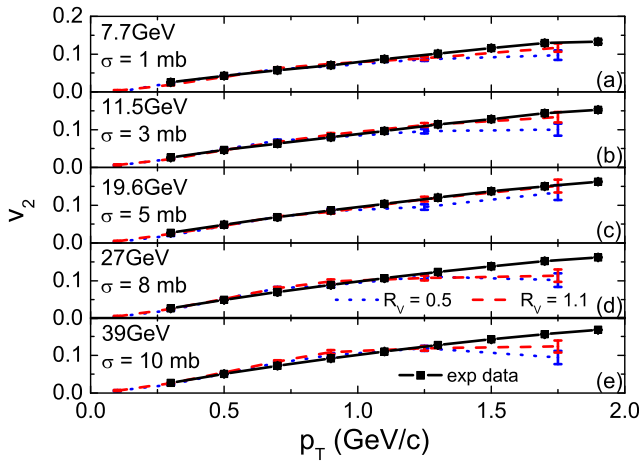}} \caption{(Color
online) Differential elliptic flow of mid-pseudorapidity
($|\eta|<1$) particles in mid-central ($20-30\%$) Au+Au collisions
at $\sqrt{s_{NN}^{}}=7.7$ (a), $11.5$ (b), $19.6$ (c), $27$ (d), and
$39$ GeV (e) with different values of $R_V$. The experimental data
are taken from Ref.~\cite{STAR12}.} \label{v2pt}
\end{figure}

We first assume that the parton scattering cross section in
the NJL transport model is isotropic and determine its value by fitting the final charged particle
elliptic flow to the experimental data. As displayed in
Fig.~\ref{v2pt}, the parton scattering cross sections of 1 mb for
7.7 GeV, 3 mb for 11.5 GeV, 5 mb for 19.6 GeV, 8 mb for 27 GeV, and
10 mb for 39 GeV can reproduce reasonably well the transverse
momentum dependence of the elliptic flow of mid-pseudorapidity
particles, by using the same sub-$|\eta|$ method as applied in the
experimental analysis~\cite{STAR12}. The reason why a larger cross
section is needed at higher energies is due to the attractive scalar
partonic potential in the NJL transport model, whose effect is more
pronounced at higher collision energies. However, the larger parton scattering cross section at higher beam energies likely leads to a decreasing specific shear viscosity $\eta/s$ of the partonic matter, i.e., the ratio of the shear viscosity $\eta$ to the entropy density $s$, with increasing temperature, contrary to results from other studies (see, e.g., Ref.~\cite{Chr15}). This is because $\eta \sim \langle p \rangle/\sigma$ with the average momentum $\langle p \rangle$ proportional to the temperature $T$ according to Ref.~\cite{Xu11b} and $s\sim T^3$ if we assume that the partonic matter consists of non-interacting massless up and down quarks, so $\eta/s\sim 1/(T^2\sigma)$ decreases with temperature. To obtain a more realistic behavior for the $\eta/s$, such as that from the NJL model~\cite{Mar13,Gho15}, requires an improved calculation using a parton scattering cross section that depends on the local temperature and density~\cite{Xu11b,Ruggieri:2013ova}. Such a study is, however, beyond the scope of the present study. Figure \ref{v2pt} further shows that reducing the strength of the
vector potential in the partonic phase by a factor of two only slightly lowers
the elliptic flow. Therefore, once the parton scattering cross section is
fitted, the relative contributions from the partonic and hadronic
phases to the elliptic flows are well constrained.

\subsection{Density evolution}

\begin{figure}[h]
\centerline{\includegraphics[scale=0.8]{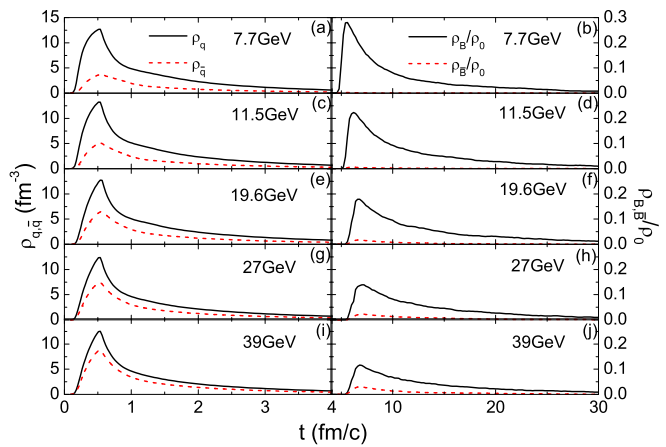}} \caption{(Color
online) Time evolution of densities of quarks and antiquarks (left
columns) as well as baryons and antibaryons (right columns) in
central cells of the partonic and the hadronic phase, respectively, in
mid-central ($20-30\%$) Au+Au collisions at $\sqrt{s_{NN}^{}}=7.7$,
$11.5$, $19.6$, $27$, and $39$ GeV.} \label{den}
\end{figure}

We display in Fig.~\ref{den} the time evolution of particle and
antiparticle densities in the baryon-rich matter produced in
mid-central ($20-30\%$) Au+Au collisions at $\sqrt{s_{NN}^{}}=7.7$,
$11.5$, $19.6$, $27$, and $39$ GeV. In the partonic phase, it is
seen that the peak quark density in central cells is similar at
different collision energies, although the lifetime of the partonic
phase is generally longer at higher collision energies. However, the
antiquark density in central cells increases with increasing beam
energy as a result of higher temperatures and smaller quark chemical
potentials reached at higher collision energies. In the hadronic
phase, the baryon density in central cells is higher at lower
collision energies compared to that at higher
collision energies, while this is the other way round for the
antibaryon density. The later appearance of hadrons at higher
collision energies is due to the later freeze-out of the partonic
phase and the additional hadron formation time of 0.7 fm/$c$
introduced in the AMPT model. It is also seen that based on the
present hadronization condition, the density in the hadronic phase
is much smaller than the saturation density $\rho_0$.

\subsection{Elliptic flow splitting}

\begin{figure}[h]
\centerline{\includegraphics[scale=0.8]{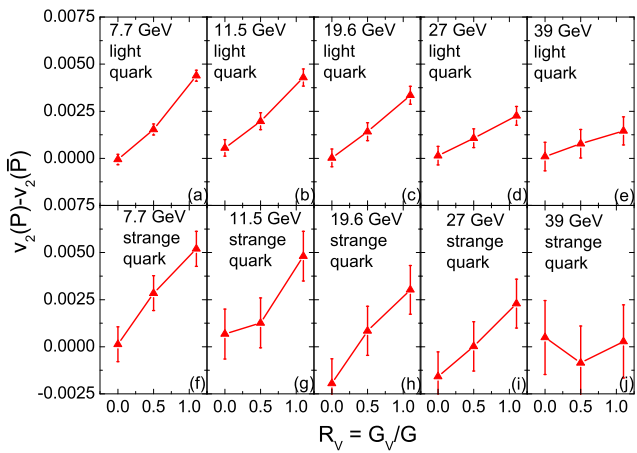}}
\caption{(Color online) The elliptic flow difference between
mid-rapidity light quarks and their antiquarks (upper panels) as
well as that between mid-rapidity strange quarks and
anti-strange quarks (lower panels) at the end of the partonic phase
in mini-bias ($0-80\%$) Au+Au collisions at $\sqrt{s_{NN}^{}}=7.7$,
$11.5$, $19.6$, $27$, and $39$ GeV, with $R_V=0$, $0.5$, and $1.1$.}
\label{v2diff_q}
\end{figure}

Figure~\ref{v2diff_q} displays the $v_2$ difference between light quarks as well as
strange quarks and their antiquarks after the partonic evolution at various collision energies. As
expected, the $v_2$ difference between quarks and their antiquarks
generally increases with increasing strength of the vector potential
denoted as $R_V$, although at higher collision energies it is not so
sensitive to $R_V$ and is much smaller. This is understandable since
the difference between the densities of quarks and antiquarks
becomes smaller at higher beam energies as shown in Fig.~\ref{den}.
In addition, the $v_2$ splitting comes mainly from the time component of the vector potential, while the space component of
the vector potential, which contributes oppositely to the $v_2$ splitting,
becomes more important at higher collision energies.

\begin{figure}[h]
\centerline{\includegraphics[scale=0.8]{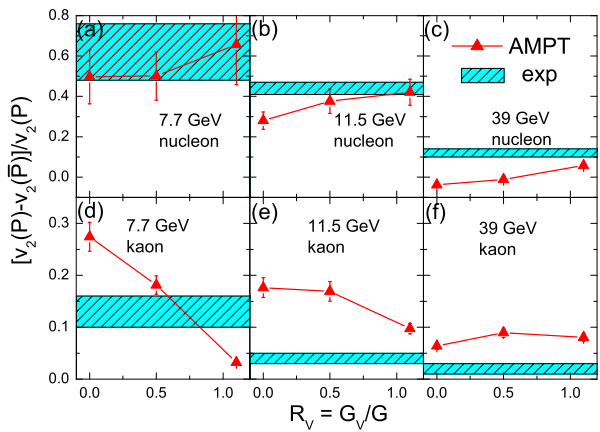}}
\caption{(Color online) The relative elliptic flow difference
between mid-rapidity nucleons and antinucleons (upper panels) and
that between mid-rapidity $K^+$ and $K^-$ (lower panels) in the
final stage of mini-bias ($0-80\%$) Au+Au collisions at
$\sqrt{s_{NN}^{}}=7.7$, $11.5$, and $39$ GeV, with $R_V=0$, $0.5$,
and $1.1$, and experimental data from Ref.~\cite{Moh11}.}
\label{v2ratio}
\end{figure}

As mentioned and
shown in Refs.~\cite{Son12,Xu14}, the initial nucleons in the hadronic
phase formed from light quarks have a larger $v_2$ than
antinucleons formed from light antiquarks. For the initial $K^+$ in
the hadronic phase, formed from a light quark and a
strange antiquark, their $v_2$ is smaller than $K^-$ formed from a
light antiquark and a strange quark, since strange quarks
(antiquarks) are heavier than light quarks (antiquarks) and
thus contribute more to the $v_2$ of produced kaons.

The relative $v_2$ difference, i.e., $v_2$ difference divided by
$v_2$ of positively charged particles, between final nucleons and
antinucleons as well as that between final $K^+$ and $K^-$ at
$\sqrt{s_{NN}^{}}=7.7$, $11.5$, and $39$ GeV are shown in
Fig.~\ref{v2ratio}. Qualitatively, the relative $v_2$ difference
between nucleons and antinucleons increases with increasing $R_V$,
while that between $K^+$ and $K^-$ mostly decreases with increasing
$R_V$. In order to reproduce the relative $v_2$ difference data from
Ref.~\cite{Moh11} at 7.7 GeV, $R_V$ is constrained between 0.5 and
1.1 by taking results from both nucleons and kaons into
consideration, consistent with the conclusion in Ref.~\cite{Xu14}.
At 11.5 GeV, it seems that an even larger value of $R_V$ is needed to
reproduce the experimental relative $v_2$ difference. At 39 GeV, the
results are, however, not so sensitive to the strength of the vector
potential, and our results underestimate the relative $v_2$
difference between nucleons and antinucleons but overestimate that
between $K^+$ and $K^-$, although the energy
dependence is qualitatively consistent with the experimental data.

\begin{figure}[h]
\centerline{\includegraphics[scale=0.8]{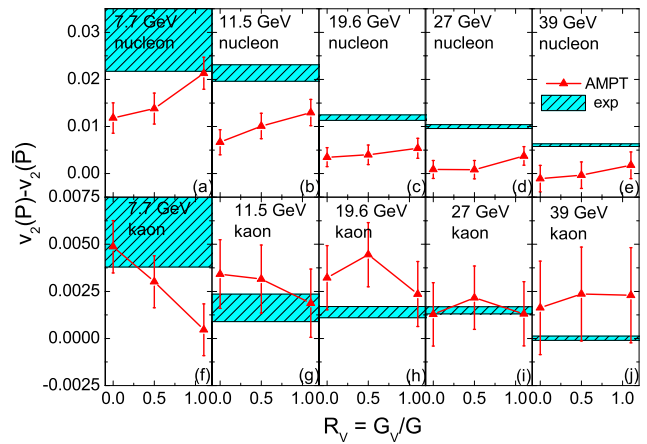}} \caption{(Color
online) The elliptic flow difference between mid-rapidity nucleons
and antinucleons (upper panels) as well as that between mid-rapidity
$K^+$ and $K^-$ (lower panels) in the final stage of mini-bias
($0-80\%$) Au+Au collisions at $\sqrt{s_{NN}^{}}=7.7$, $11.5$,
$19.6$, $27$, and $39$ GeV, with $R_V=0$, $0.5$, and $1.1$, and
experimental data from Ref.~\cite{STAR13}. } \label{v2diff}
\end{figure}

In Fig.~\ref{v2diff}, the absolute $v_2$ difference between nucleons and antinucleons as
well as $K^+$ and $K^-$ are compared with experimental results from
Ref.~\cite{STAR13} at various collision energies. The absolute $v_2$ difference raises more
challenges to the model than the relative $v_2$ difference, since this requires to reproduce the $v_2$ of various particle species as well. At lower collision energies, the $v_2$ difference
between nucleons and antinucleons increases with increasing $R_V$,
and that between $K^+$ and $K^-$ decreases with increasing $R_V$. At
7.7 GeV, the $v_2$ difference between nucleons and antinucleons
favors $R_V=1.1$, while that between $K^+$ and $K^-$ favors values
of $R_V$ between 0 and 0.5. At higher collision energies, although
our model gives qualitatively the correct sign as well as the energy
dependence of the $v_2$ splitting, it underestimates the $v_2$
difference between nucleons and antinucleons but reproduces that
between $K^+$ and $K^-$ within the statistical error.

\section{Conclusions and outlook}\label{summary}

Based on an extended AMPT model, which includes
mean-field potentials in both the partonic and the hadronic phase, uses the mix-event coalescence for hadronization,
and ensures the charge conservation during the hadronic evolution, we have studied the
energy dependence of elliptic flow splitting between particles and
their antiparticles at RHIC-BES energies. The density evolutions of
particles and their antiparticles in both the partonic and hadronic
phases are illustrated. The elliptic flow
splitting from the contribution of the partonic phase and the
further modification in the hadronic phase at various collision energies is
observed. Our model can describe reasonably well the elliptic flow
splitting at lower collision energies, and can describe
qualitatively but not quantitatively that at higher beam energies.
Especially, our model underestimates the elliptic flow splitting
between nucleons and antinucleons at higher collision energies. The
present study thus calls for other mechanisms in addition to the
mean-field potentials that may contribute to the elliptic flow
splitting between particles and their antiparticles.

The present model can be further improved in several ways for a
better description of the collision dynamics at RHIC-BES energies. First, the
yield ratio of baryon/antibaryon from the present AMPT model is
different from that obtained experimentally, or that based on the
baryon chemical potential and temperature at chemical freeze-out
from fitting the experimental data using the
thermodynamical model. This could be improved by modifying
the initial parton species or the coalescence algorithm~\cite{ZWL}.
Second, the mixing and interaction between the partonic phase and
the hadronic phase are still missing in our model, but they could be
important in heavy-ion collisions at RHIC-BES energies. Third, the
annihilation process for baryons and antibaryons could be
overestimated in the model, as the elliptic flow difference between nucleons
and antinucleons can be increased by reducing the
annihilation effect. Fourth, it is of interest to include the isovector coupling~\cite{Liu16}
in the NJL transport model and the
symmetry energy effect in the hadronic phase,
as this would allow us to study, respectively, the elliptic flow difference between $\pi^+$ and $\pi^-$ and
the interesting isospin dynamics in heavy ion collisions at the RHIC-BES and FAIR-CBM
energies.

\section*{Acknowledgments}

We thank Nu Xu for the suggestion of carrying out this study, Chen Zhong for
maintaining the high-quality performance of the computer facility, Lie-Wen Chen for helpful comments, and
Lilin Zhu for the hospitality at Sichuan University.
The work of JX was supported by the Major State Basic Research
Development Program (973 Program) of China under Contract Nos.
2015CB856904 and 2014CB845401, the National Natural Science
Foundation of China under Grant Nos. 11475243 and 11421505, the
"100-talent plan" of Shanghai Institute of Applied Physics under
Grant Nos. Y290061011 and Y526011011 from the Chinese Academy of
Sciences, the Shanghai Key Laboratory of Particle Physics and
Cosmology under Grant No. 15DZ2272100, and the "Shanghai Pujiang
Program" under Grant No. 13PJ1410600, while that of CMK was supported by the US Department of Energy under Contract No. DE-SC0015266
and the Welch Foundation under Grant No. A-1358.

\end{document}